\documentclass[aps,prd,amsmath,twocolumn,floatfix]{revtex4}

\usepackage{graphicx}
\usepackage{dcolumn}
\usepackage{bm}
\usepackage{verbatim}
\usepackage{amssymb}
\usepackage{epstopdf}
\begin{document}

\preprint{APS/123-QED}

\title{Spherical Collapse and Cluster Counts in Modified Gravity Models}

\author{Matthew C. Martino, Hans F. Stabenau \& Ravi K. Sheth}
\email{mcmarti2@sas.upenn.edu, shethrk@physics.upenn.edu}
\affiliation{Department of Physics and Astronomy, University of Pennsylvania, Philadelphia, PA 19103}

\label{firstpage}

\begin{abstract}
Modifications to the gravitational potential affect the nonlinear gravitational evolution of large scale structures in the Universe.  To illustrate some generic features of such changes, we study the evolution of spherically symmetric perturbations when the modification is of Yukawa type; this is non-trivial, because we should not and do not assume that Birkhoff's theorem applies.  We then show how to estimate the abundance of virialized objects in such models.  Comparison with numerical simulations shows reasonable agreement:  When normalized to have the same fluctuations at early times, weaker large scale gravity produces fewer massive halos.  However, the opposite can be true for models that are normalized to have the same linear theory power spectrum today, so the abundance of rich clusters potentially places interesting constraints on such models.  Our analysis also indicates that the formation histories and abundances of sufficiently low mass objects are unchanged from standard gravity.  This explains why simulations have found that the nonlinear power-spectrum at large $k$ is unaffected by such modifications to the gravitational potential.  In addition, the most massive objects in CMB-normalized models with weaker gravity are expected to be similar to the high-redshift progenitors of the most massive objects in models with stronger gravity.  Thus, the difference between the cluster and field galaxy populations is expected to be larger in models with stronger large-scale gravity.  
\end{abstract}

\maketitle

\section{Introduction}
When we consider the cosmological data from WMAP, supernovae Ia, galaxy clustering on large scales, and cross-correlations between galaxies and the CMB, we are faced with several possible conclusions. One is that we live in a spatially flat Friedmann-Robertson-Walker universe currently dominated by either a cosmological constant or repulsive dark energy. The best fit for the dimensionless energy density parameters are $\Omega_m=0.28$ and $\Omega_\Lambda=0.72$, within the concordance $\Lambda$CDM Model \cite{wmap5}. The advantage of this model is that nearly all available observational data supports it; the disadvantage is that it requires the vast majority of the energy density in the universe to be in two unknown substances, dark matter and dark energy.  

This conclusion rests on the accuracy of our current gravity model, General Relativity.  The key equation of General Relativity, Einstein's equation, relates the curvature and the expansion rate of the Universe to its matter and energy content. The current paradigm is to modify the matter content of the universe, by including dark matter and dark energy, to account for observations. Instead, however, we might modify how the universe curves in response to matter, which would mean modifying our theory of gravity.

There are many possible ways to modify gravity, depending on what one wishes to ``fix''. For example, MOND \cite{milgrom, bekenstein} removes the need for dark matter to account for galactic rotation curves and has several other interesting results, but seems to fail at the scale of galaxy clusters, with even optimistic accountings needing roughly as much dark matter as baryonic matter. At the other end is something like DGP or conformal gravity, which hopes to account for the acceleration of the universe without invoking a cosmological constant \cite{dgp, lss, mannheim, diaferio}.  In addition, there are other models which seek to unify dark matter and dark energy \cite{chaplygin, makler, scherrer05, scherrer08, bertacca}. What we seek to do in this paper is to study the problem more phenomologically. For example, regardless of how the force law for gravity is modified, it will often be stronger or weaker, relative to the standard model, at larger or shorter scales.  One way to parameterize this is to introduce a modified Yukawa-like potential for a point mass:
\begin{equation}
 \label{initialpotential}
 \phi(r)=G m \frac{1+\alpha (1-e^{-r/r_s})}{r},
\end{equation}
\cite{sealfon,shirata1,shirata2,fritz} where $\alpha$ indicates the strength and $r_s$ the scale (constant in physical rather than comoving coordinates) on which this modification is most relevant.  On scales smaller than $r_s$, $\phi(r)$ reduces to the standard Newtonian potential; on larger scales it transitions to the Newtonian potential multiplied by a factor of $(1+\alpha)$. This is similar to the interaction considered in \cite{ngp}, in which a long range dark matter interaction is introduced yielding a different Yukawa-like potential that instead modifies gravity on short length scales.

Note that this is not a cosmological model:  there is no prescription for determining things like the expansion factor and the resulting Hubble factor.  Like previous workers, we will assume that these are the same as in the standard cosmology.  The main goal of studying such a model is to gain intuition for some generic effects of modifications to standard gravity.  

For example, in the linear theory of such a model, the growth of fluctuations is $k$-dependent \cite{shirata1} -- it is not in standard gravity.  As a result, a smooth spherical region within which the density is the same as the background universe will evolve.  This qualitatively different behavior from standard gravity has not been emphasized -- so it is worth showing the argument explicitly.  Consider the density field smoothed on scale $R$ at some early time $t_i$.  We can write this field in terms of its Fourier modes, and the (Fourier Transform of the) smoothing kernel as 
\begin{equation}
 \delta_R(\mathbf{x},t_i) = \int d\mathbf{k}\,
      \exp(i\mathbf{k}\cdot\mathbf{x})\,\delta(\mathbf{k})\, W(kR).
\end{equation}
The linearly evolved field is 
\begin{equation}
 \delta_R(\mathbf{x},t) = \int d\mathbf{k}\,
      \exp(i\mathbf{k}\cdot\mathbf{x})\,
        \frac{D(k,t)}{D(k,t_i)}\,\delta(\mathbf{k})\, W(kR),
\end{equation}
where $D$ is the linear theory growth factor.  In standard gravity, $D$ is independent of $k$, so if $\delta_R(\mathbf{x},t_i)=0$ then $\delta_R(\mathbf{x},t)=0$ also.  But if $D$ depends on $k$, then if $\delta_R(\mathbf{x})=0$ at some time $t$, it will, in general, be non-zero at other times (the exception being if the $k$-dependence of $W$ happens to exactly cancels that of $D$).  For the Yukawa-like modification considered here, the $k$ dependence of the spherical top hat filter does not cancel that of $D$.  Thus, we are led to the rather remarkable conclusion that, when the gravitational potential has been modified, then linear theory predicts that a spherical tophat patch within which the density is the same as the background will evolve!  The reason why can be traced to the fact that Birkhoff's Theorem no longer applies once the Newtonian potential has been modified.  Without this Theorem, the spherical top hat filter is no longer special, and our common sense prejudice from standard gravity -- that initially overdense regions become denser, underdense regions less dense, but regions within which the density is the same as the background do not evolve -- must be treated with caution.  


The evolution of nonlinear clustering in this model has been studied using numerical simulations by \cite{fritz,shirata2}.  Our goal in what follows is to provide a framework for understanding this nonlinear evolution more completely.  To this end, we will use the spherical evolution model, which has found extensive use in the standard model -- it is used to motivate estimates of the abundance of nonlinear objects \cite{pressSchechter}, a crucial ingredient in methods which describe the growth of nonlinear gravitational clustering \cite{cooraySheth}.  It also provides a framework for discussing the environmental dependence of clustering \cite{moWhite,st02}.  

Section~\ref{model} summarizes what is known from the linear theory of this Yukawa-like model, and then shows our estimate of the key, and sometimes subtle, changes to the spherical evolution model.  Section~\ref{sims}  compares this work to the simulations of \cite{fritz}. A final section summarizes.  

\section{The model}\label{model}

\subsection{Linear theory and a slightly modified potential}

We begin by considering the evolution of density perturbations.  This can be done by either considering the fluid equations in expanding coordinates, or by considering the conservation of stress-energy $\nabla_aT^{ab}=0$. If we start with a smooth background, add small perturbations, and linearize the resulting equations, we get a second order differential equation for the evolution of the fractional overdensity, $\delta(x,t)$ that depends on time, scale factor $a$, the Hubble parameter $H=\dot a/a$, and the potential $\phi$. In standard gravity, we would use the Poisson equation to set the relationship between $\phi$ and $\delta$, but here we will assume a modified Poisson equation that results in the above potential, equation~(\ref{initialpotential}).  
\begin{equation}
 \label{linearrealspace}
 \ddot{\delta}+2 H(t) \dot{\delta} = \nabla^2 \phi
\end{equation}
It is easier to work with the Fourier transform of this equation:
\begin{equation}
 \label{eq:linear}
 \ddot{\delta}_k+2 H \dot{\delta}_k = \frac{3}{2}
        \left[1+\alpha \frac{a^2}{a^2+k^2r_s^2}\right] 
        \left(1-\frac{H_0^2 \Omega_\Lambda}{H^2} \right) H^2 \delta_k.
\end{equation}
This can be solved relatively easily to determine the linear growth of structure associated with equation~(\ref{initialpotential}).  

\begin{figure}[tp] 
   \centering
   \includegraphics[width=0.70\hsize]{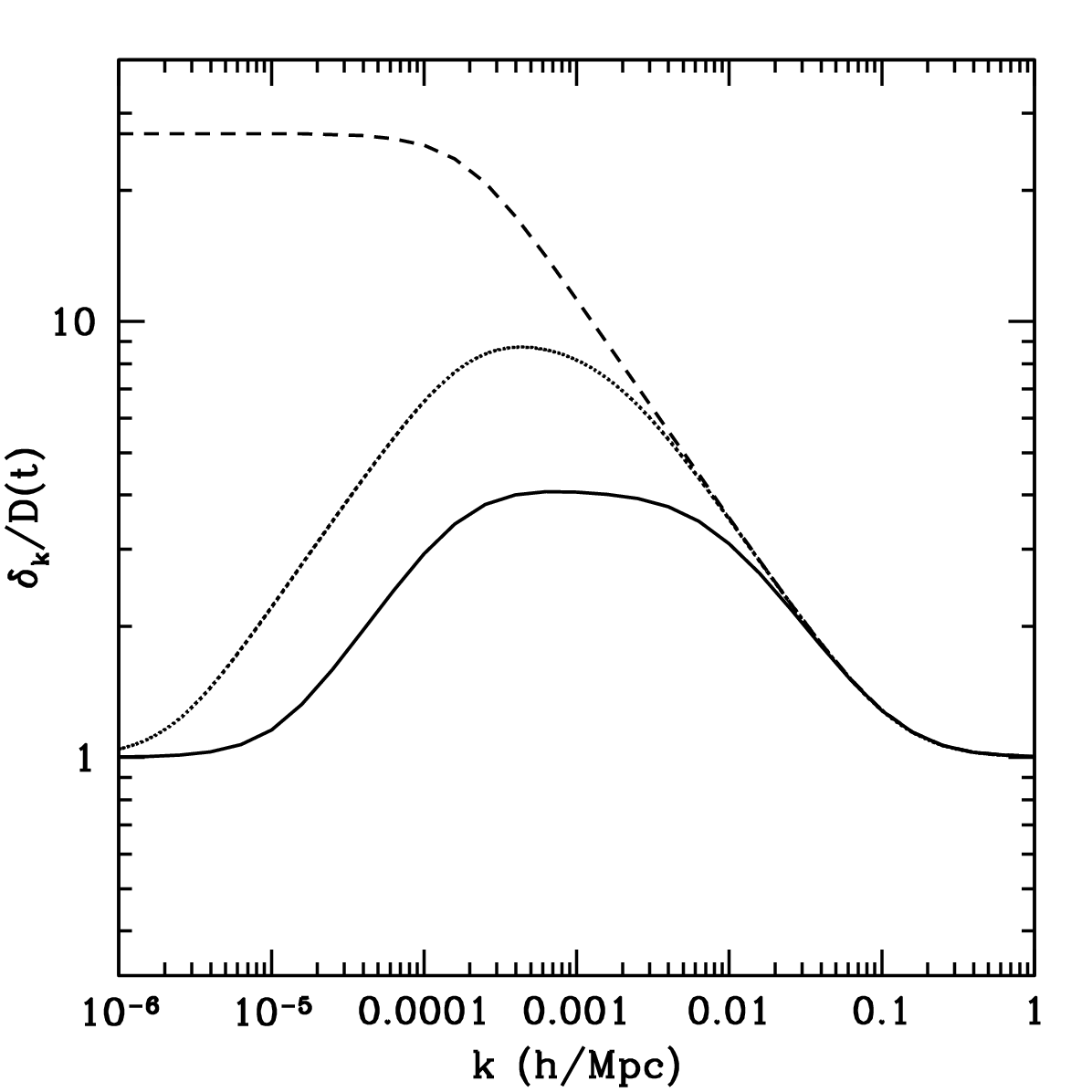}
   \includegraphics[width=0.70\hsize]{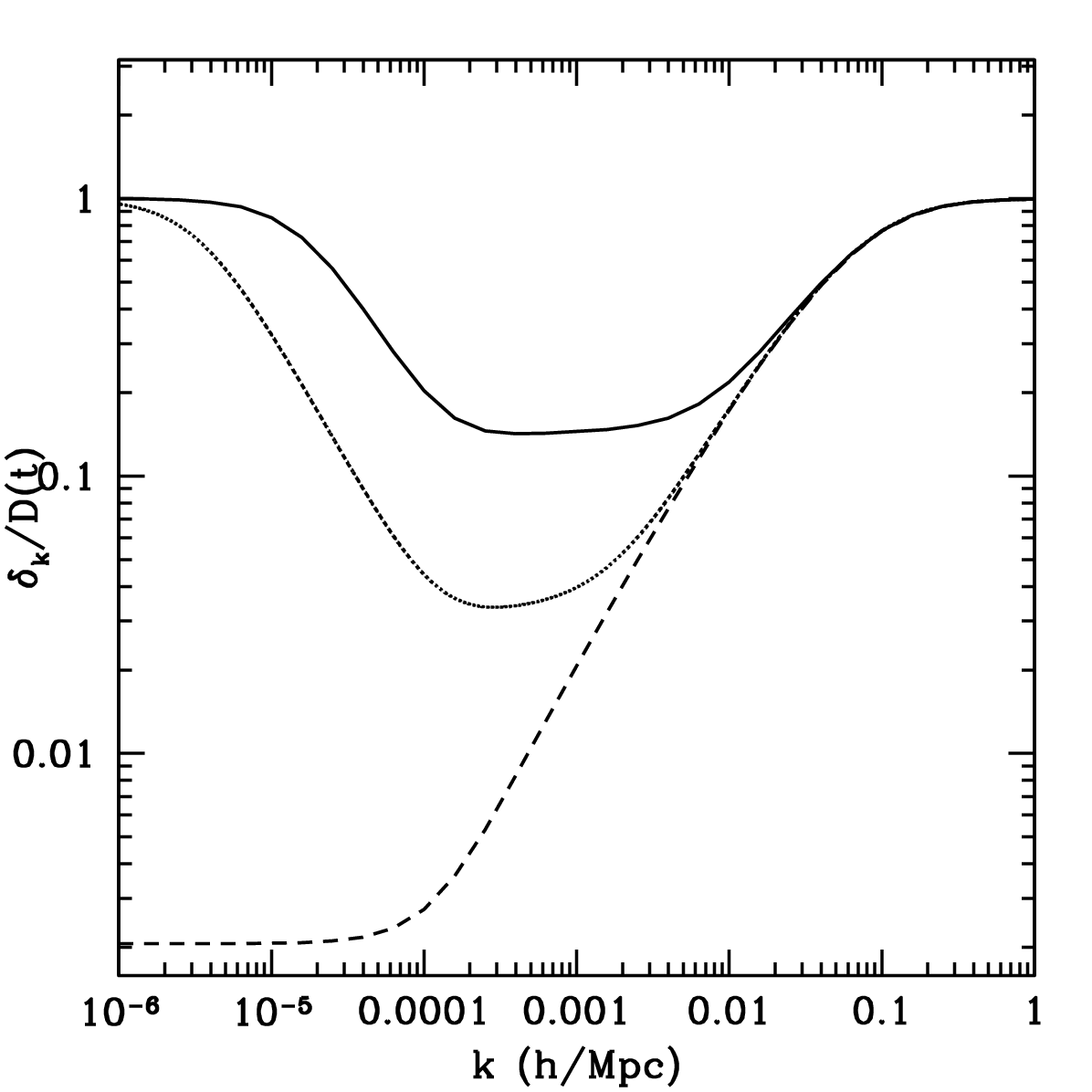}
   \caption{Ratio of linear theory growth factor to that in standard $\Lambda$CDM, at $a=1$, when $r_s=5h^{-1}$~Mpc and $\alpha=1$ (top) and $\alpha=-1$ (bottom).  Dashed lines show this ratio for the model in equation~(\ref{initialpotential}), and solid lines for equation~(\ref{potentialeq}).  For the solid lines, $r_c=70h^{-1}$~Mpc, and dotted $r_c=350h^{-1}$~Mpc.}
   \label{fig:linear}
\end{figure}

The dashed lines in Figure~\ref{fig:linear} show how this growth differs from that in the standard model.  Note in particular that $\delta_k^{\rm Lin}(t) = D(k,t)\,\delta_k^{\rm initial}$, whereas the standard model has no $k$ dependence in the growth factor $D(t)$.  The figure shows the effects that one expects to see, namely that for negative $\alpha$ the growth factor is smaller at small $k$ (large scales), whereas for positive $\alpha$ the growth factor is larger at large scales. Both have a region where they deviate from being scale independent, until on very large scales they return to scale independence, though with a different value than in General Relativity.

While this is not a significant problem for linear theory, we decided to explicitly force the potential back to normal gravity at large scales for reasons which will become clear in the next section, briefly, because we want to assume that the cosmological model is indistinguishable from $\Lambda$CDM, on the largest scales we want gravity to be the same as in $\Lambda$CDM. The potential we are using from this point onward is:
\begin{equation}
\label{potentialeq}
\phi(r)=G m \frac{1+\alpha (1-e^{-r/r_s}) - \alpha (1-e^{-r/r_c})}{r}
\end{equation}
With this potential, the $r_c\gg r_s$ term serves to make explicit the return to normal gravity at large scales. The linear theory equation becomes 
\begin{eqnarray}
\label{linear}
 \ddot{\delta}_k+2 H \dot{\delta}_k &=& 
  \frac{3}{2}\left[1+\alpha \frac{a^2}{a^2+k^2r_s^2} - \alpha \frac{a^2}{a^2+k^2r_c^2} \right] \nonumber\\
 &&\qquad \times \left(1-\frac{H_0^2 \Omega_\Lambda}{H^2} \right) H^2 \delta_k,
\end{eqnarray}
and the linear growth associated with this solution is shown as the solid lines in Figure~\ref{fig:linear}, when $r_c=70h^{-1}$~Mpc.  The dotted line shows what happens if $r_c=350h^{-1}$~Mpc -- a scale which is large compared to that probed by BAOs.  Although the analysis which follows uses $r_c=70h^{-1}$~Mpc, the results which follow are not sensitive to this choice.

\subsection{Spherical collapse}

Excursion set methods \cite{bondetal, laceyCole, rks98} are used to estimate the number density of collapsed haloes, the merger rates of haloes, the conditional mass function of progenitors as a function of final halo mass and time, and the nonlinear counts-in-cells distribution, all of which help us to link what we observe about the properties of galaxies and galaxy clusters with cosmology. Essentially, excursion set methods relate the properties of halos today to the initial density fluctuation field.  An advantage of such methods is that we only need a few things in order to use them:  one is an assumption that the initial fluctuations are small and Gaussian, the other is a model for determining how dense something must have been initially to collapse at a given time. The first is given to us by WMAP, the second is more difficult.

A simple model for how to determine this critical density is given by the spherical evolution model, in which one considers a spherical tophat perturbation in the initial density fluctuation field.  In the standard model, the gravitational evolution of such a patch is determined only by the mass within it, and so one can determine how overdense such a patch needs to be initially in order to collapse by a given time.  This critical overdensity $\delta_c$ generically depends on the background cosmology \cite{gunnGott}.

The spherical collapse calculation begins with the statement that the force driving the acceleration is related to the gravitational potential by
\begin{equation}
 \label{F=gradPhi}
 \frac{d^2 r}{d t^2} = F = -\nabla \Phi(r).
\end{equation}
This can be integrated once to get
\begin{equation}
 \label{solveeq}
 \frac{1}{2} \left( \frac{d r}{d t} \right )^2 +\Phi(r) = C,
\end{equation} 
where $\Phi(r)$ is the integral of the potential over the mass distribution, and $C$ is the total energy of the patch, which is constant.  In standard gravity, the potential of a shell of mass $M$ is the same as that of a point mass at the center of the sphere, so $\Phi(r)$ reduces to $GM(<r)/r$.  The constant $C$ can be related to the initial overdensity and/or expansion rate of the patch:  the initial expansion rate is given by the Hubble expansion rate of the background in which the patch is embedded, namely in comoving coordinates $\dot x_i=0$, so $\dot r_i=\dot a_i x_i=\dot a_i/a_i (a_i x_i)$, so $(dr/dt)_i=H_ir_i$.  Including a cosmological constant presents no conceptual difference.  

In standard gravity one can directly solve this equation.  The solution is a cycloid for which the critical value of the initial overdensity required for collapse, $\delta_c$, does not depend on the initial size of the patch.  This scale independence of $\delta_c$ is a result of Birkhoff's Theorem:  the evolution of a tophat sphere is the same as that given by Friedmann's equations, so the actual size of the patch drops out. 

When gravity is modified, things are no longer so simple.  For example, when the potential is given by equation~(\ref{potentialeq}), Birkhoff's theorem no longer applies:  A particle offcenter in a uniform spherical shell will feel a force from the shell because we no longer have a $1/r^2$ force law.  This has two consequences.  First, equation~(\ref{F=gradPhi}) can still be integrated once to get equation~(\ref{solveeq}), only now $\Phi(r)$ has contributions from both the internal and external mass distributions.  We can get $\Phi$ by integrating equation~(\ref{potentialeq}) of the mass distribution, leaving $C$ and the initial value for $dr/dt$ to be determined.  As before, $C$ is the total energy (constant in time), and we set $(dr/dt)_i=H_ir_i$.  (This was why we used equation~\ref{potentialeq} rather than equation~\ref{initialpotential}.)  Second, whereas evenly spaced concentric shells remain evenly spaced in the standard tophat model, this is no longer the case when the potential is modified.  As a result, the initial tophat perturbation develops a nontrivial density profile as it collapses.

\begin{figure}[tp]
  \centering
  \includegraphics[width=0.70\hsize]{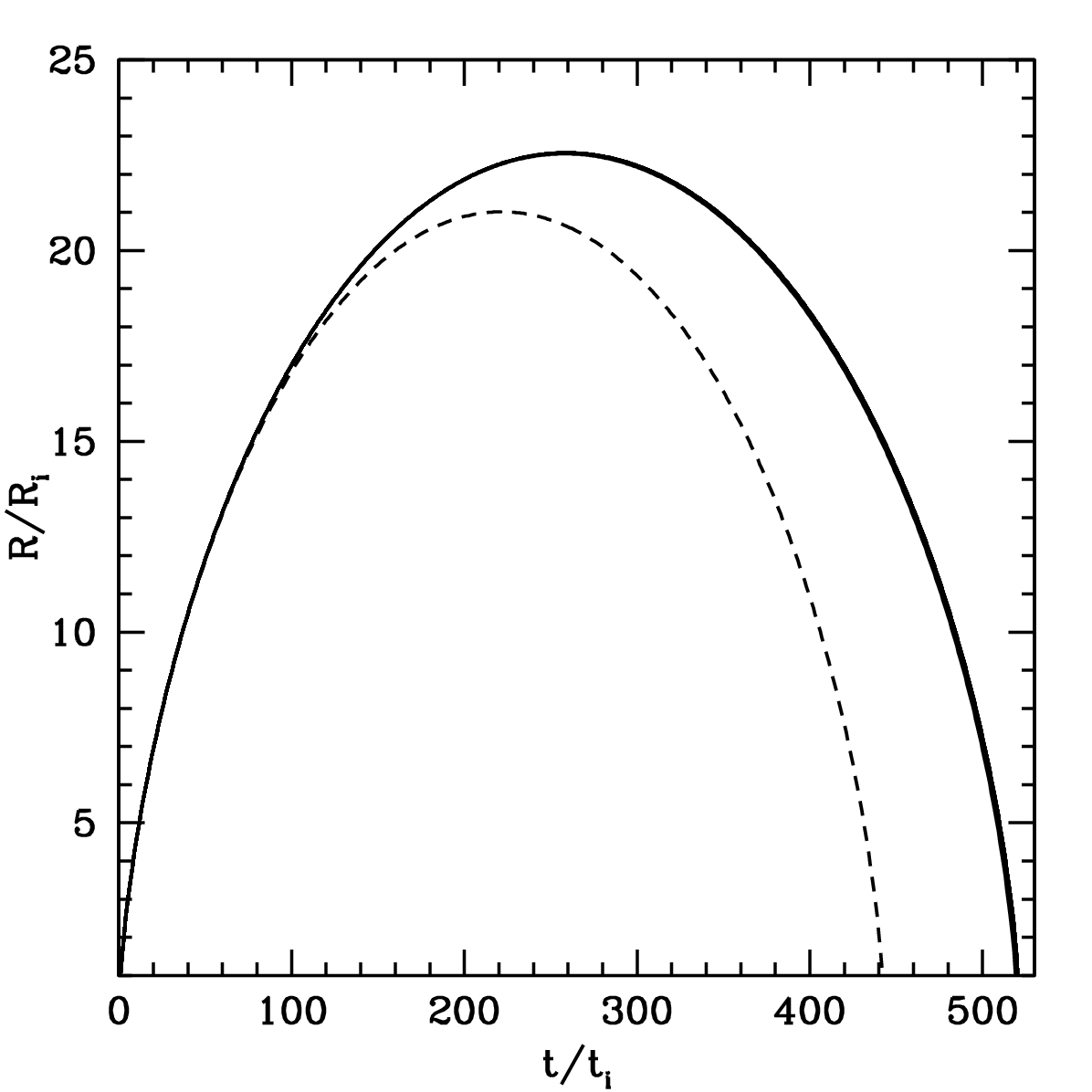}
  \includegraphics[width=0.70\hsize]{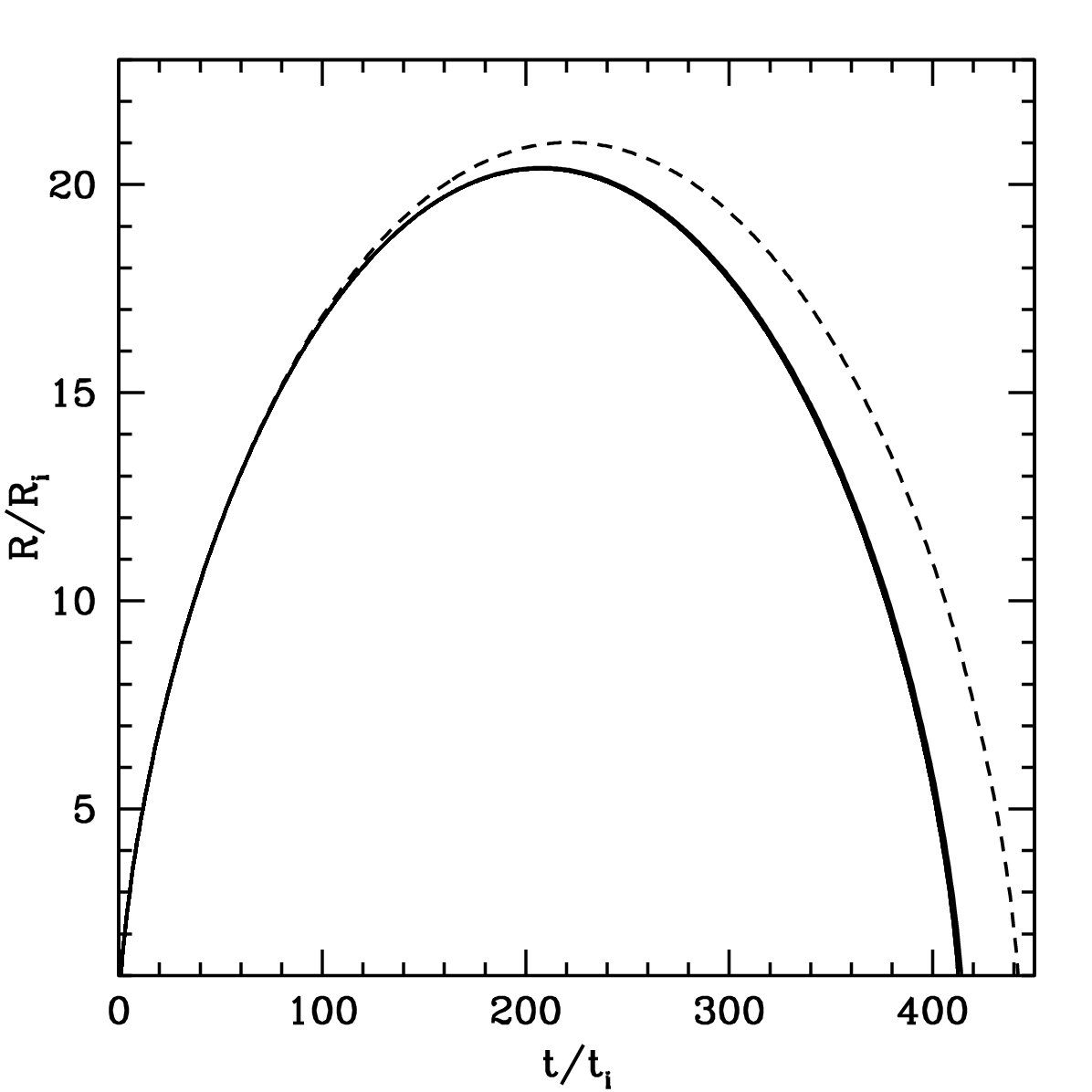}
  \caption{Convergence to solution as number of shells increases for two objects with mass $10^{14.5}h^{-1}M_\odot$, with $\alpha=0.5$ (top) and $\alpha-1.0$ (bottom).  The thick solid lines are actually the positions of the shell on the edge of the density perturbation for  3,10, 20, 40, 50, 60, 80, 100, and 200 shells, note they all lie nearly on top of each other. The dashed curves show the evolution when $\alpha=0$. }
  \label{fig:denplot}
 \end{figure}

Of course, neither of these changes prevent us from solving for the evolution of $r(t)$.  Our main interest in what follows is not in the details of how the density profile is modified (this is interesting in its own right), but in the modification to the critical density required for collapse.  To estimate this in practice, two things require care, both of which are related to the breakdown of Birkhoff's theorem.  The first is that, because the shape of the perturbation evolves, one must follow the evolution not just of a shell at the boundary of the perturbation, but of a series of concentric shells.  So the question is:  How finely spaced must the shells be before one converges to the correct solution?  The second is that one now cares not only about the mass initially interior to the initial boundary, but the mass exterior as well.  In this case, the question is: How far beyond the initial boundary of the perturbation matters before one reaches convergence?  

Therefore we start with an initial patch which is substantially larger than that within which there is an initial overdensity, and use a simple 1-dimensional N-body simulation to solve for $r(t)$.  We found that volumes having twice the initial comoving radius or larger were sufficient to account for the lack of Birkhoff's theorem, regardless of $\alpha$ or $r_s$.  We also found that for objects of mass up to $10^15h^{-1}M_\odot$, using 3 (linearly spaced) wide shells provided a $\delta_c$ that was within 1\% of 200 shells (see Figure~\ref{fig:denplot}). For higher mass objects we found that we needed more than 3 shells, but by 40 shells $\delta_c$ is within 0.02\% of $\delta_c$ calculated with 200 shells.


\begin{figure}[tp] 
   \centering
   \includegraphics[width=0.95\hsize]{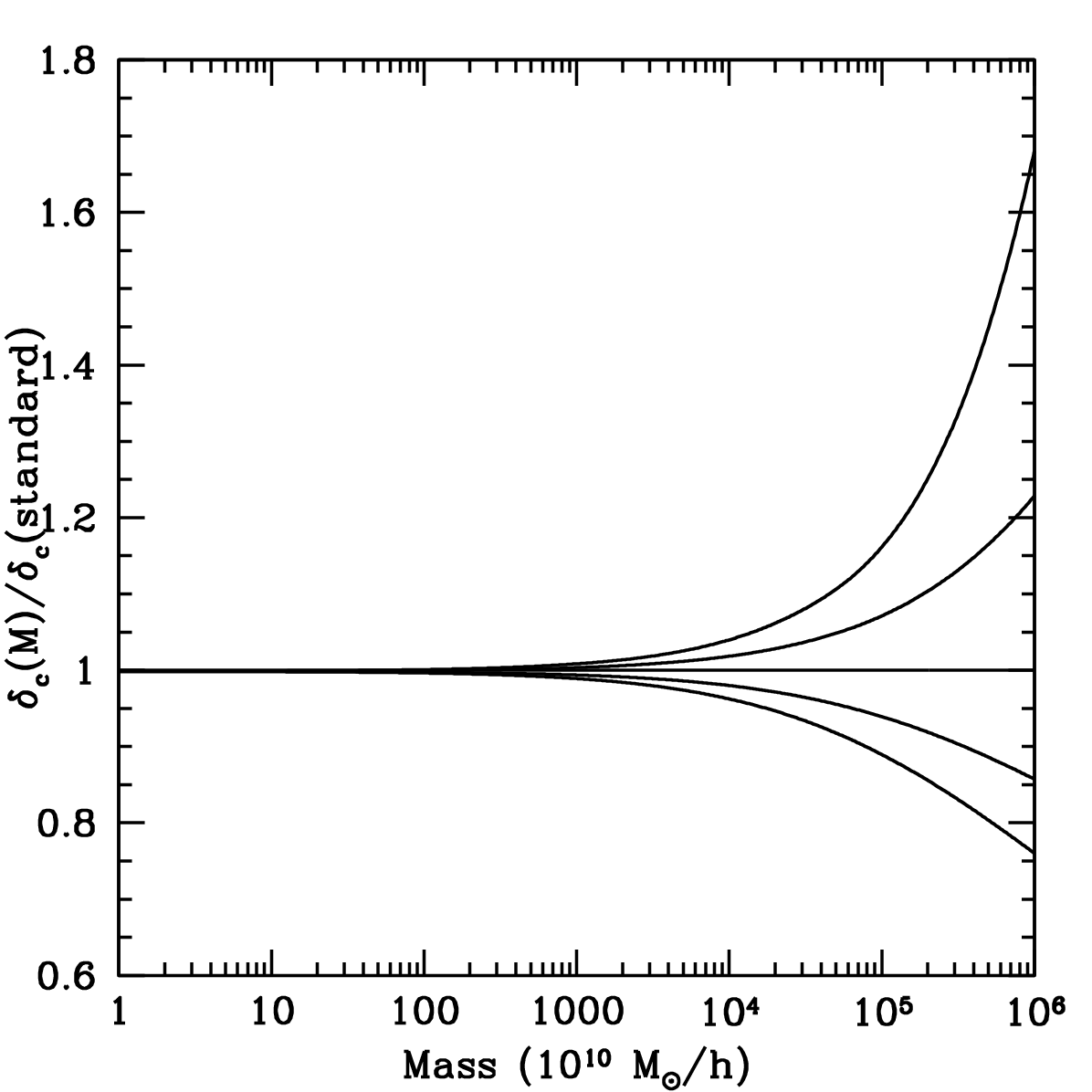}
   \caption{Ratio of initial density required for collapse at the present time to that in the standard gravity model, when the background cosmology is $\Lambda$CDM, $r_s=5h^{-1}$ Mpc and $r_c=70h^{-1}$ Mpc.  From top to bottom, curves show models in which $\alpha=-1, -0.5, 0, 0.5$ and $1$ (note that $\alpha=0$ is standard gravity). }
   \label{fig:barriershapes}
\end{figure}

Having determined that our numerical solution had converged, we evaluated $r(t)$ at the present time in an $\Lambda$CDM background model, for a grid of initial sizes and overdensities.  By finding which pairs of initial density and size $r_i$ when evolved result in $r(t_0)=0$, we obtained $\delta_{ci}(r_i)$.  Since the initial overdensity is always small, we can use the fact that 
 $M = (4 \pi /3) r_i^3 \bar\rho_i(1+\delta_i)\approx (4 \pi /3) r_i^3$ 
to express this critical density as a function of mass rather than initial radius.  This is shown in Figure~\ref{fig:barriershapes}.  Notice that $\delta_c$ depends on mass; this is not unexpected, because patches which remain smaller than $r_s$ throughout their evolution (and become small mass halos) are unlikely to notice any modification, whereas those which are larger than $r_s$ at any time during their evolution will.  This mass-dependence of $\delta_c$ means that we expect to see a variation in cluster abundances only at masses larger than $\sim 10^{14}h^{-1}\,M_\odot$.  This is a consequence of the fact that $r_s = 5h^{-1}$~Mpc, for which the mass scale is about $5\times 10^{13}h^{-1}\,M_\odot$.  (For fixed $\alpha$, the mass scale is proportional to $r_s^3$.)  

\begin{figure}[tp] 
   \centering
   \includegraphics[width=0.95\hsize]{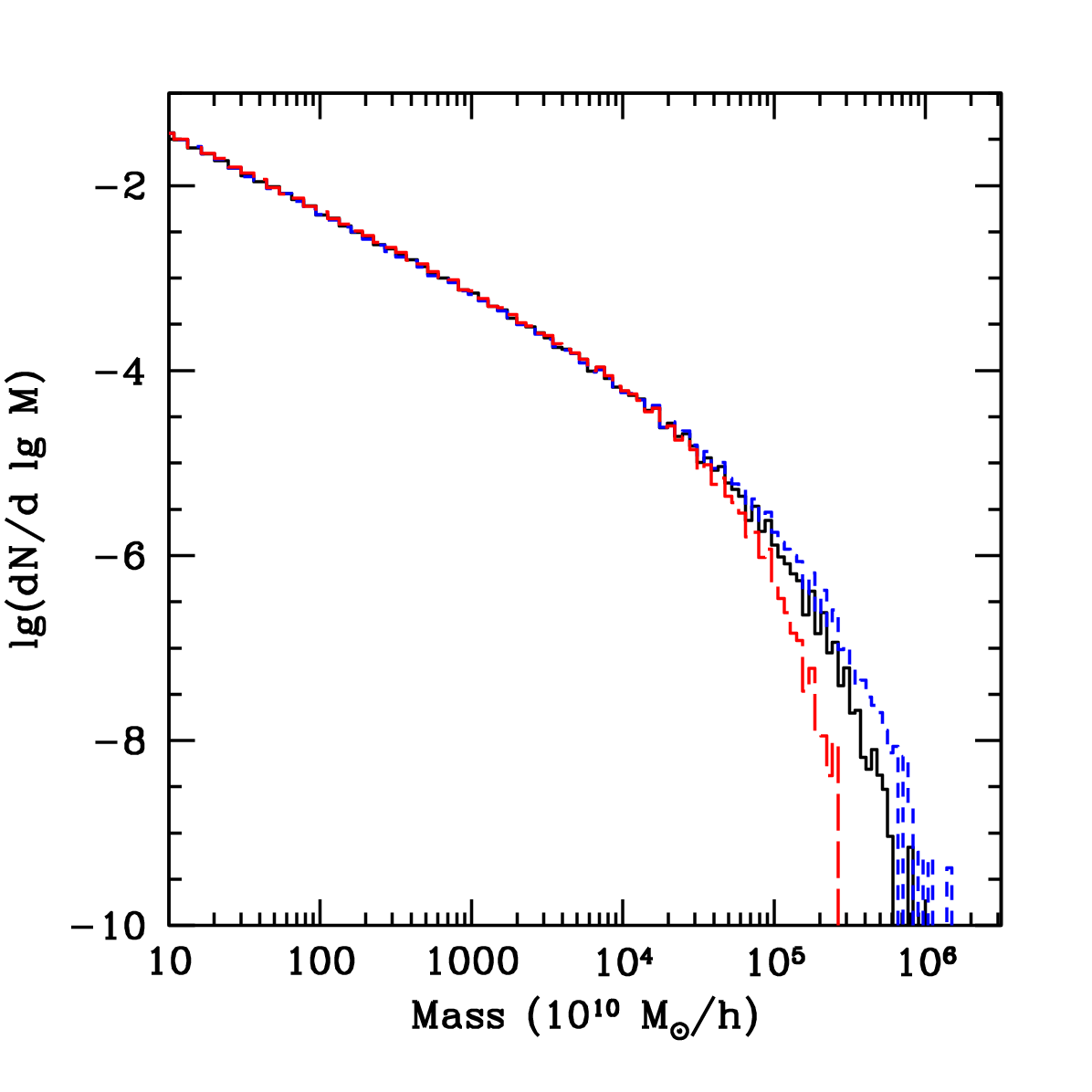}
   \caption{(Color online) Expected halo abundance, $\log dN/d\log m$, as a function of mass $m$ for models with $\alpha=1$ (short dashed) and $\alpha=-1$ (long dashed), the solid line is for $\alpha=0$. The rapid rise of the negative $\alpha$ barrier in figure~\ref{fig:barriershapes} is why there is a larger effect on the abundance of high mass halos for negative $\alpha$ than positive $\alpha$. }
   \label{fig:comparison}
\end{figure}

\subsection{The abundance of virialized objects}
Figure~\ref{fig:barriershapes} shows that, when the potential is modified, then $\delta_c$ is no longer scale-independent.  Because it depends on mass, the relevant excursion set problem is one with a `moving' rather than `constant' barrier, so it is of the type first studied by \cite{st02}.  Figure~\ref{fig:comparison} shows the result of using this formalism to estimate the abundance of virialized objects.  Briefly, making this estimate requires that one generate an ensemble of random walks in the $(\delta_i,S_i)$ plane, where $S_i\equiv \sigma^2_i(M)$ is the variance in the initial fluctuation field when smoothed on scale $r_i$.  Since $\sigma_i(M)$ is a monotonic function of $M$, the variables $S_i$, $M$ and $r_i$ are essentially equivalent to one another.  In particular, 
\begin{equation}
 S_i \equiv \int \frac{dk}{k}\,\frac{k^3 P_i(k)}{2\pi^2}\, W^2(kr_i),
\end{equation}  
where $W(x) = (3/x^3)\,(\sin x - x \cos x)$.  
One then finds the first crossing distribution $f(S_i)dS_i$ of the `barrier' $\delta_{ci}(M) = \delta_{ci}(S_i)$.  The abundance of objects is $dn/d\ln M\,d\ln M = (\bar\rho/M)\,f(S_i)dS_i$.  

Figure~\ref{fig:comparison} shows that the abundance of massive objects increases as $\alpha$ increases, whereas the abundance of low mass objects is essentially unchanged from when $\alpha=0$.  We argued above that this makes intuitive sense:  the lower mass objects do not feel the change in gravity because they were smaller than $r_s$ throughout their evolution; the more massive objects are able to become even more massive if $\alpha$ is positive, since then gravity is stronger.

\subsection{Choice of normalization and incompatibility with standard gravity}\label{standardcomp}
Before moving on, it is worth noting that we were careful to describe and implement the excursion set approach in initial rather than linearly evolved variables.  In standard gravity ($\alpha=0$), it is more common to phrase the discussion in terms of linearly evolved variables.  Since the linear growth factor is independent of $k$ when $\alpha=0$, this is straightforward.  However, this is no longer the case when $\alpha\ne 0$, because of the $k$-dependence in $D(k,t)$.

Indeed, the barrier shape in Figure~\ref{fig:barriershapes} is qualitatively like that of the linear growth factor in Figure~\ref{fig:linear}, so one might ask if the difference that we see in the mass function is entirely a consequence of the $k$-dependence of the linear growth factor.  More specifically, the differences shown in Figure~\ref{fig:comparison} are really a consequence of two effects:  first, we have assumed that $S_i(M)$ is the same for all $\alpha$.  Therefore, $S_0(M)$ is not:  the linear theory evolution when $\alpha>0$ results in more large scale power than when $\alpha\le 0$, so the rms fluctuation in the present day fields are different -- in the jargon, $\sigma_8$ at $z=0$ is larger for the $\alpha>0$ models.  Since we know that, in standard gravity, the abundance of massive halos is exponentially sensitive to $\sigma_8$, one might wonder if this alone accounts for much of the effect.  (Later on, we will discuss another consequence of normalizing the models at the initial rather than final time.)  To quantify this, we would like to compare the predicted abundances when the models are normalized so that $S_0(M)$, rather than $S_i(M)$, is the same.  This will isolate the effect of a mass-dependent $\delta_c(M|\alpha,r_s,r_c)$ on the halo abundances.  

\begin{figure}[tp] 
   \centering
   \includegraphics[width=0.70\hsize]{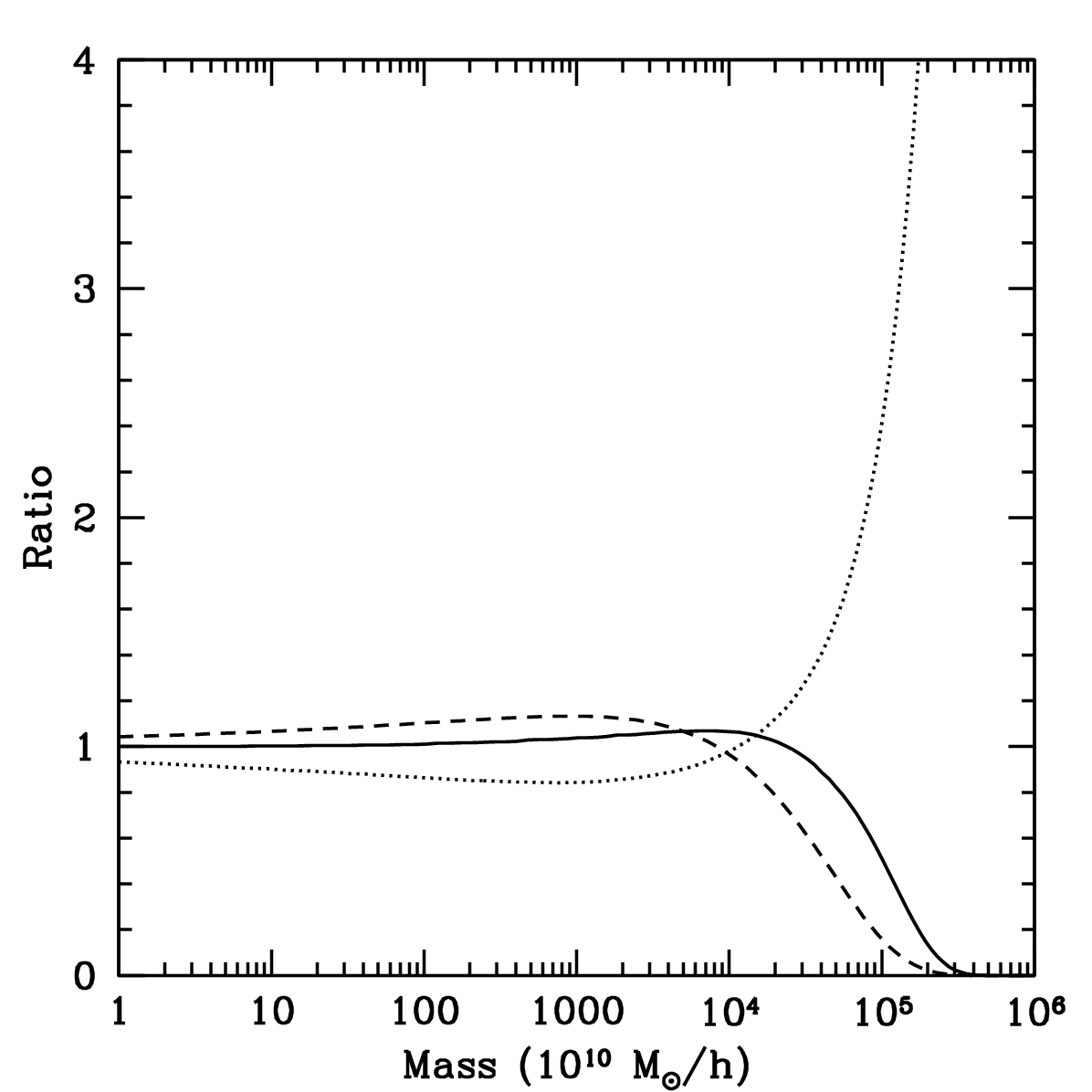}
   \includegraphics[width=0.70\hsize]{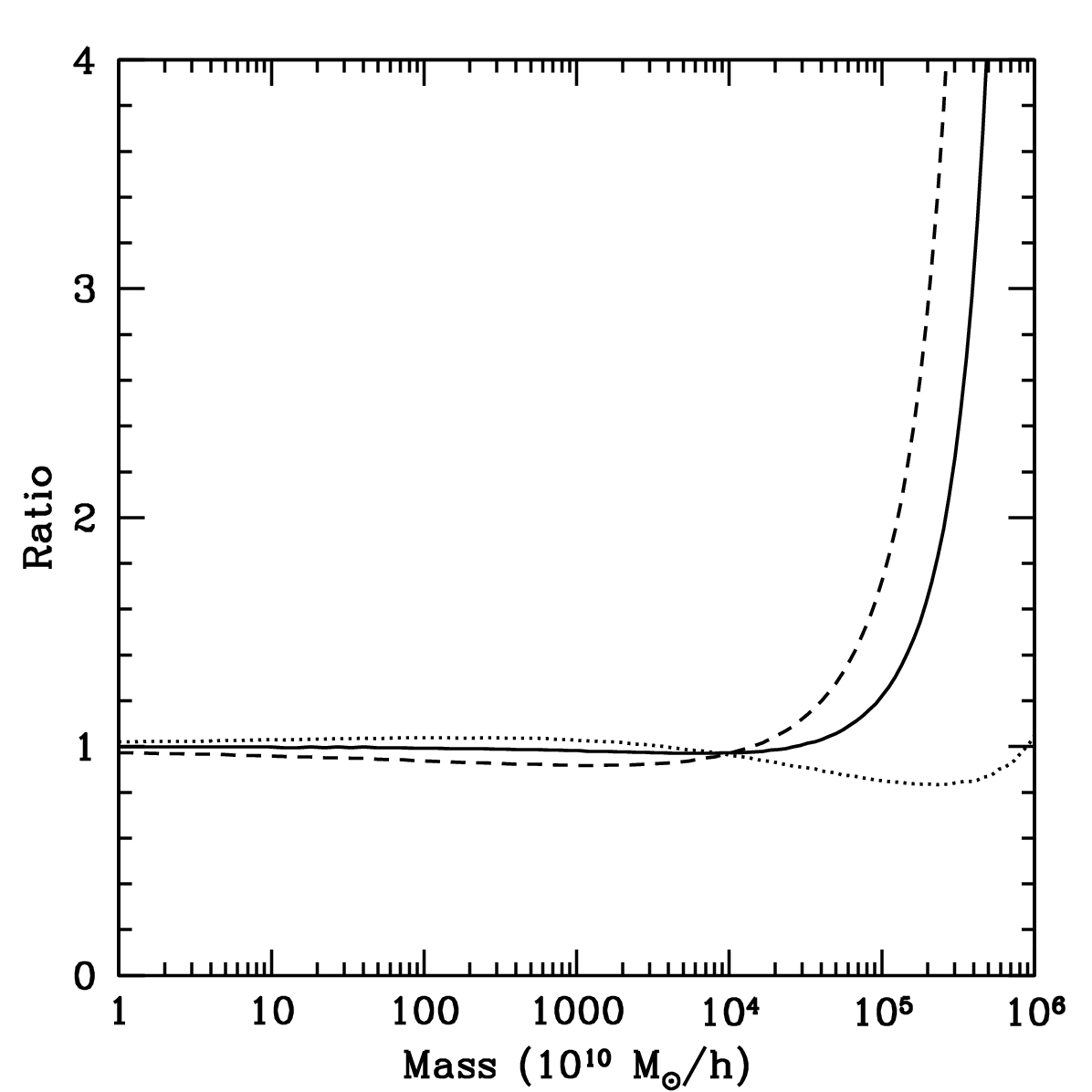}
   \caption{Ratio of halo abundances when $\alpha=-1$ (top) and $\alpha=0.5$ (bottom) to that when $\alpha=0$. Solid curve shows this ratio when the excursion set method is correctly implemented, using the initial fluctuation field values $S_i(M)$ and a moving barrier $\delta_c(M)$, and $S_i(M)$ is independent of $\alpha$.  Dashed curve uses $S_0(M)$ from the linearly evolved field and a constant barrier $\delta_c=1.686$.  Dotted curve uses $S_i(M)$ and $\delta_c(M)$, but now $S_i(M)$ is modified so that $S_0(M)$ is the same for both values of $\alpha$. Note that all these approaches produce different results. }
   \label{fig:comparisonlinear}
\end{figure}

Therefore, we used the excursion set approach with the same constant barrier height as one would have in the standard ($\alpha=0$) linearly evolved gravity model, and then evaluated $S_0(M)$, rather than $S_i(M)$, using the $k$-dependent linear growth factor.  I.e., 
\begin{equation}
 S_0 \equiv \int \frac{dk}{k}\,D^2(k,t_0)\,
                 \frac{k^3 P_i(k)}{2\pi^2}\, W^2(kr_i),
\end{equation}  
The resulting first crossing distribution $f(S_0)dS_0$ is the same as in standard gravity (after all the barrier is constant), but when expressed as a function of mass $M$, the abundances are different because the relation between $S_0$ and $M$ depends on $(\alpha,r_s,r_c)$.  The dashed line in Figure~\ref{fig:comparisonlinear} shows that this method yields a mass function that also differs substantially from the standard one:  it drops substantially below unity for large $M$.  In this case, however, the drop is entirely due to the $k$-dependent growth factor.  In addition, notice that although it is qualitatively like the solid line, for which $\delta_c$ is mass dependent, it can be substantially different (i.e., the ratio of the solid to the dashed line is greater than unity) at large $M$.  This shows that there is more to the change in the mass function than simply the change in the relation between $S$ and mass.  

Whereas the solid line shows results in which the initial power spectra are the same for all $\alpha$ (i.e., $S_i(M)$ is the same for all models), the dashed line shows what happens if we adjust the shape of the initial power spectrum in the $\alpha=0$ model so that $S_0(M)$ is the same as for the $\alpha=-1$ model.  The dotted line shows the result of adjusting instead the initial $P(k)$ of the $\alpha=-1$ model so that it produces the same $S_0(M)$ as the original $\alpha=0$ calculation.  In this case, the $\alpha=-1$ initial conditions now have substantially more large scale power, so the predicted abundance of massive halos is larger, until the mass dependence of $\delta_c$ begins to matter (this is not evident in the figure, because it happens at $M >10^{16.7}~ M_\odot/h$).  

The fact that none of the curves shown in Figure~\ref{fig:comparisonlinear} are unity for all $M$, nor are any two curves the same, means that cluster abundances in modified gravity models cannot be mimicked in standard gravity simply by changing the shape of the initial power spectrum so that it agrees with modified linear theory at $z=0$.  Trading `CMB'-normalization for `cluster' normalization does not work, because the cluster mass function depends on the nonlinear physics of gravitational collapse:  Cluster counts are sensitive to more than the change to linear theory.  For CMB-normalized models, we feel that the appropriate estimate of the effect of modifying gravity is shown by the solid line.  In the following section we use numerical simulations to test this prediction.  

\begin{figure}[tp] 
   \centering
   \includegraphics[width=0.475\hsize]{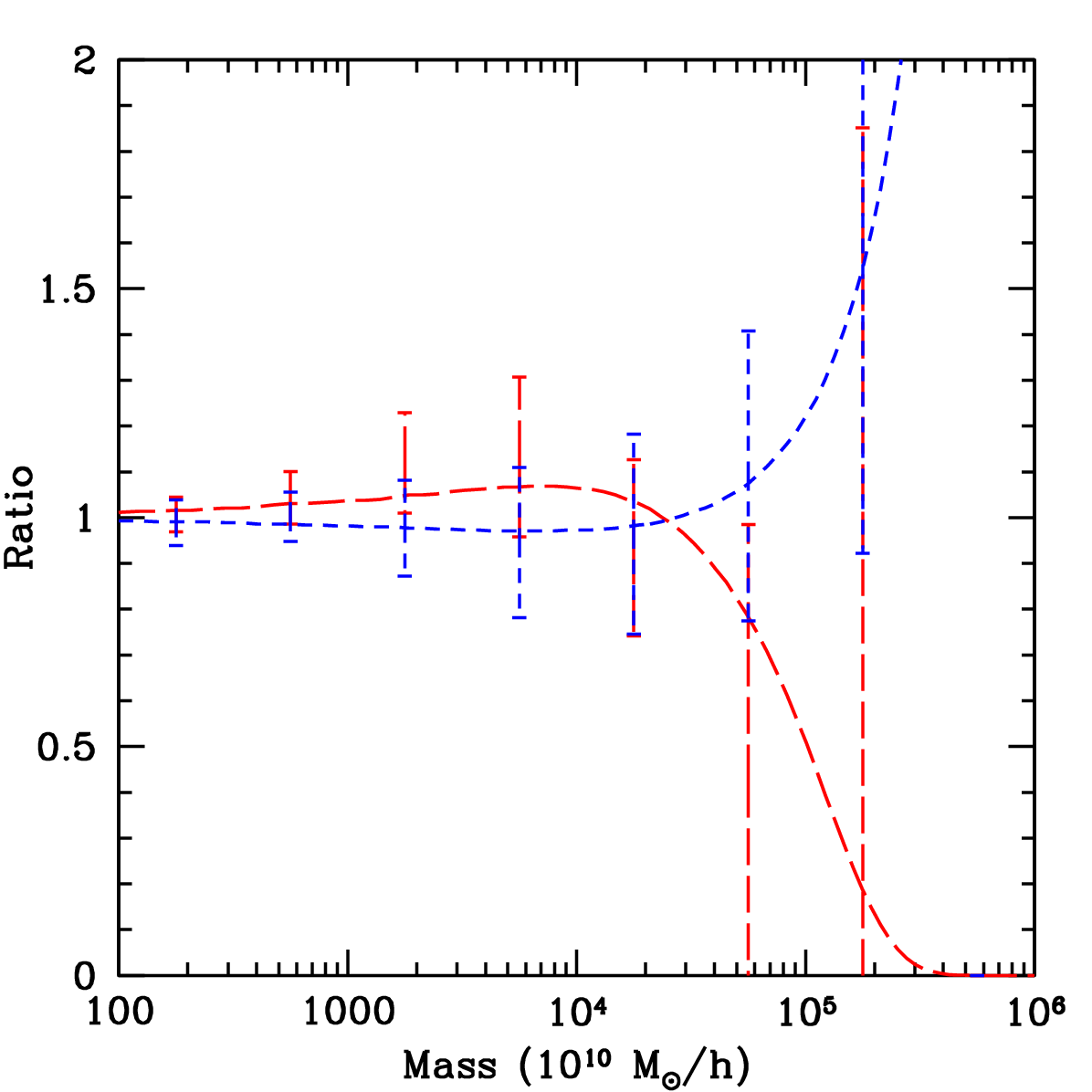}
   \includegraphics[width=0.475\hsize]{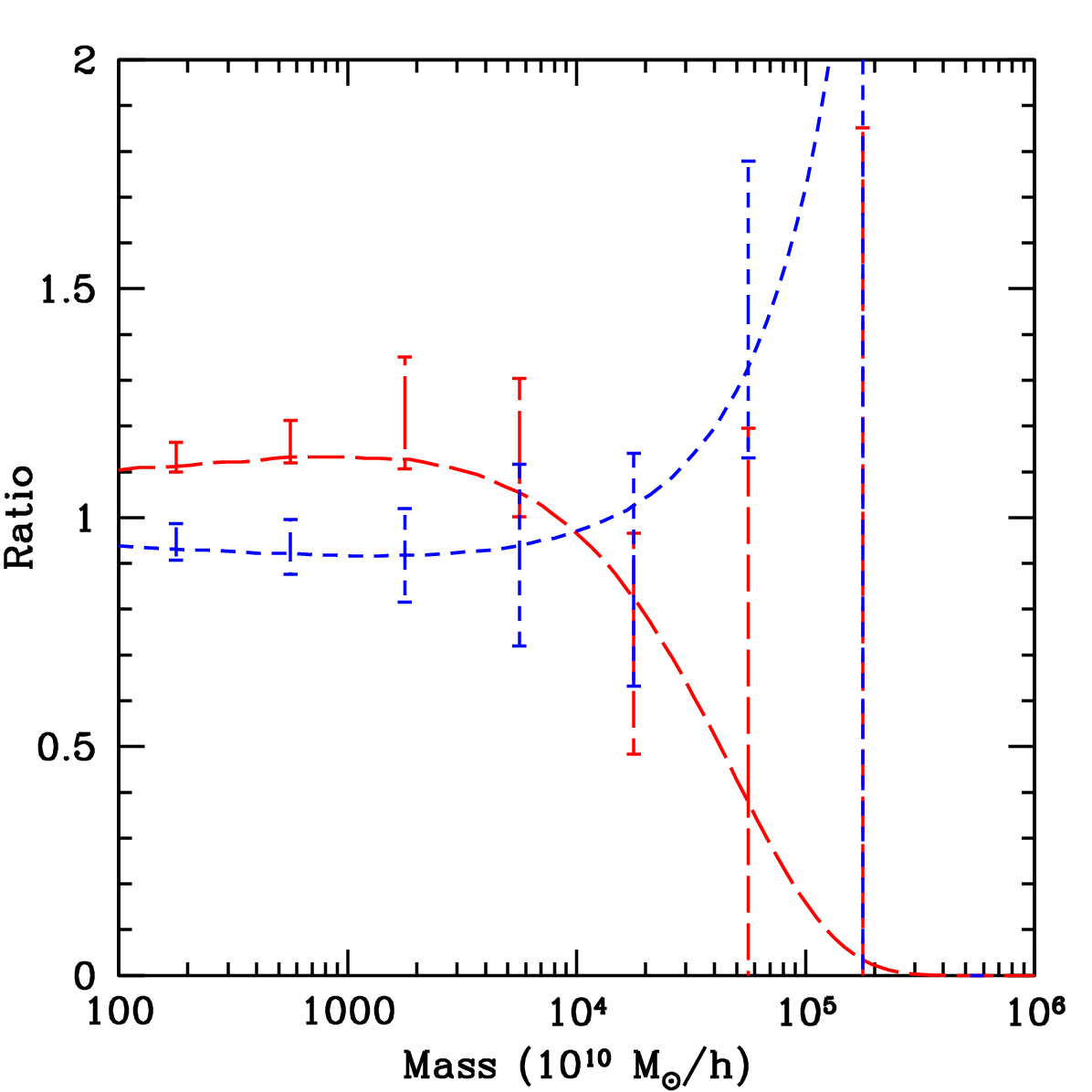}
   \includegraphics[width=0.475\hsize]{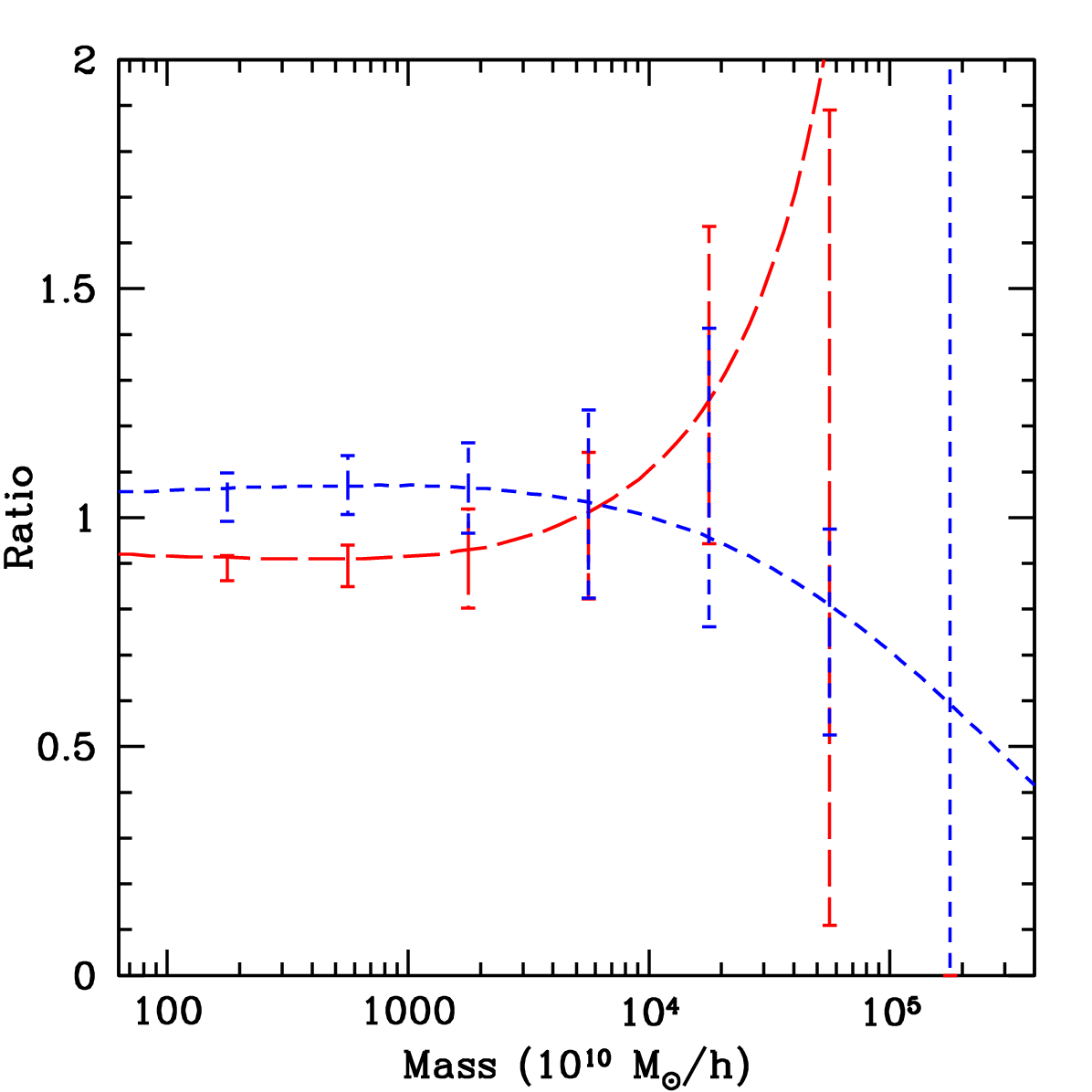}
   \caption{(Color online) Halo abundances with $r_s=5$~Mpc ratioed to $\alpha=0$.  In all panels, data points are from simulation, whereas curves are from our excursion set calculation.  In the top left panel, (blue) long dashed and (red) short dashed curves are for $\alpha=0.5$ and $-1$, respectively, and all models had the same initial fluctuation spectrum (so that $\alpha=0$ has $\sigma_8=1$ today).  The top right panel shows the ratio of counts in two $\alpha=0$ runs, but with different initial conditions, taylored so that $\sigma_8$ at $z=0$ is the same as in the $\alpha=0.5$ (long dashed) and $-1$ (short dashed) runs.  Short dashed curves in bottom panel show the ratio of the $\alpha=0.5$ counts to that in the $\alpha=0$ run when it has been normalized to have the same (linear theory) power spectrum at $z=0$ as the $\alpha=0.5$ run.  Long dashed curve shows a similar analysis of the $\alpha=-1$ case.  This panel shows the ratio of the numerators in the previous two panels. }
    \label{fig:compareall}
\end{figure}

\section{Comparison to simulations}\label{sims}
We now compare our spherical collapse predictions for halo abundances with measurements in the simulations of \cite{fritz}.  These simulations followed the evolution of $128^3$ particles in a periodic box of size $100 h^{-1}$Mpc, for various choices of $\alpha$ and $r_s$.  In all cases, the background cosmology was flat $\Lambda$CDM with $\Omega=0.3$, and the particle mass was $1.1\,\times\,10^{10} M_\odot$.  In addition, the simulations were always started from the same initial phases, a feature we will exploit shortly.  We identify halos in the standard way using a friends-of-friends algorithm with link length 0.2 times the interparticle separation.  In what follows, we show results from the $r_s=5h^{-1}$~Mpc runs. The $\alpha=0$ simulation, with standard initial conditions has $\sigma_8=1.0$ at $z=0$, the corresponding runs for $\alpha=0.5$ and $\alpha=-1$ have $\sigma_8=1.2$ and $0.7$ respectively.  Following our discussion of how the counts depend on the shape and normalization of the initial power spectrum, we also study results from $\alpha=0$ simulations in which the initial power spectrum was modified so that, at $z=0$, it has the same shape as the two $\alpha\ne 0$ cases ($\sigma_8=1.2$ and 0.7).

Figure~\ref{fig:compareall} shows how the mass function depends on $\alpha$.  The panels shows the ratio to $\alpha=0$, and curves show the predictions from our excursion set with moving or standard barrier calculation with standard or modified initial power spectrum.  The calculation is in reasonably good agreement -- note in particular that it captures the sense of the trends with $\alpha$.

\begin{figure}[tp]
   \centering
   \includegraphics[width=0.70\hsize]{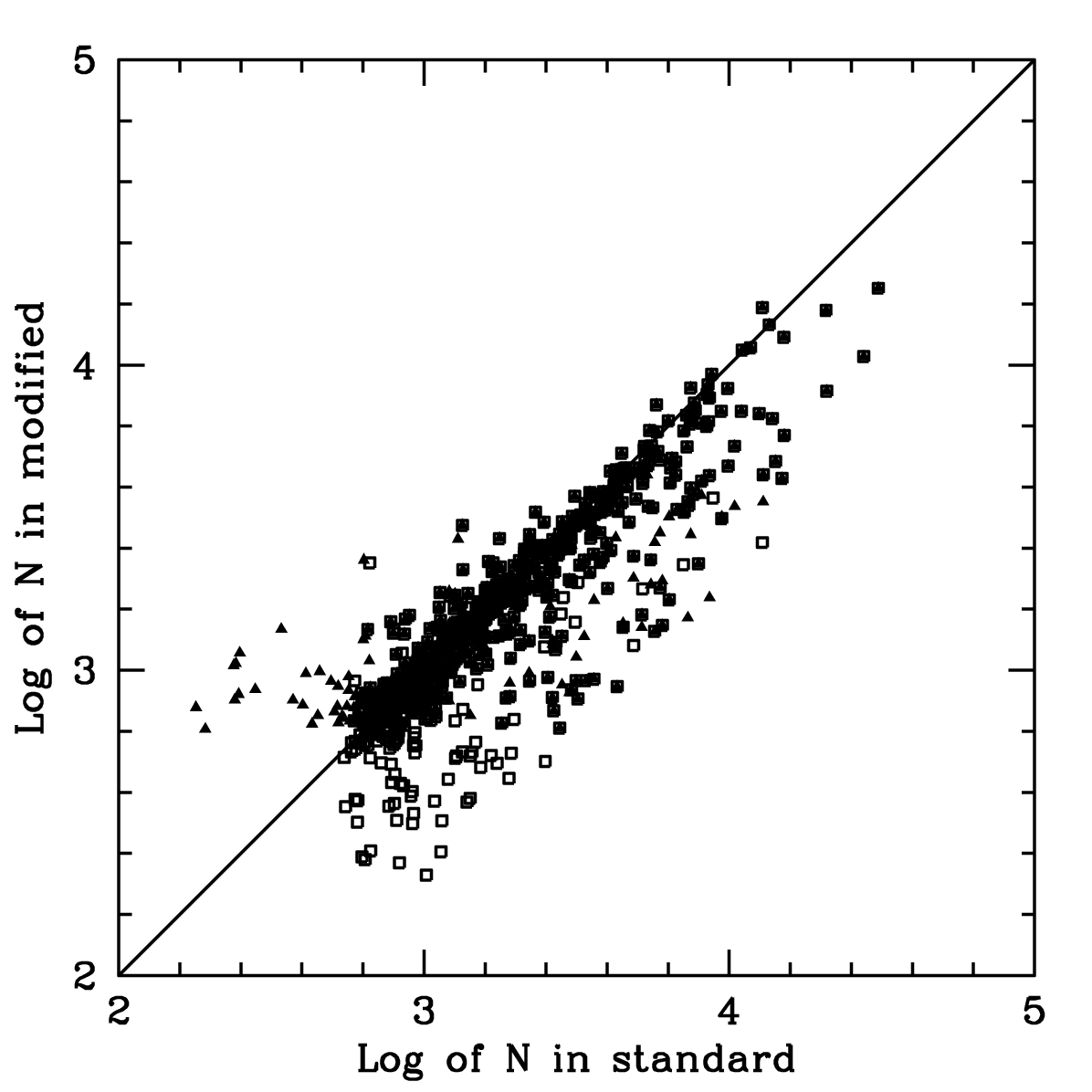}
   \includegraphics[width=0.70\hsize]{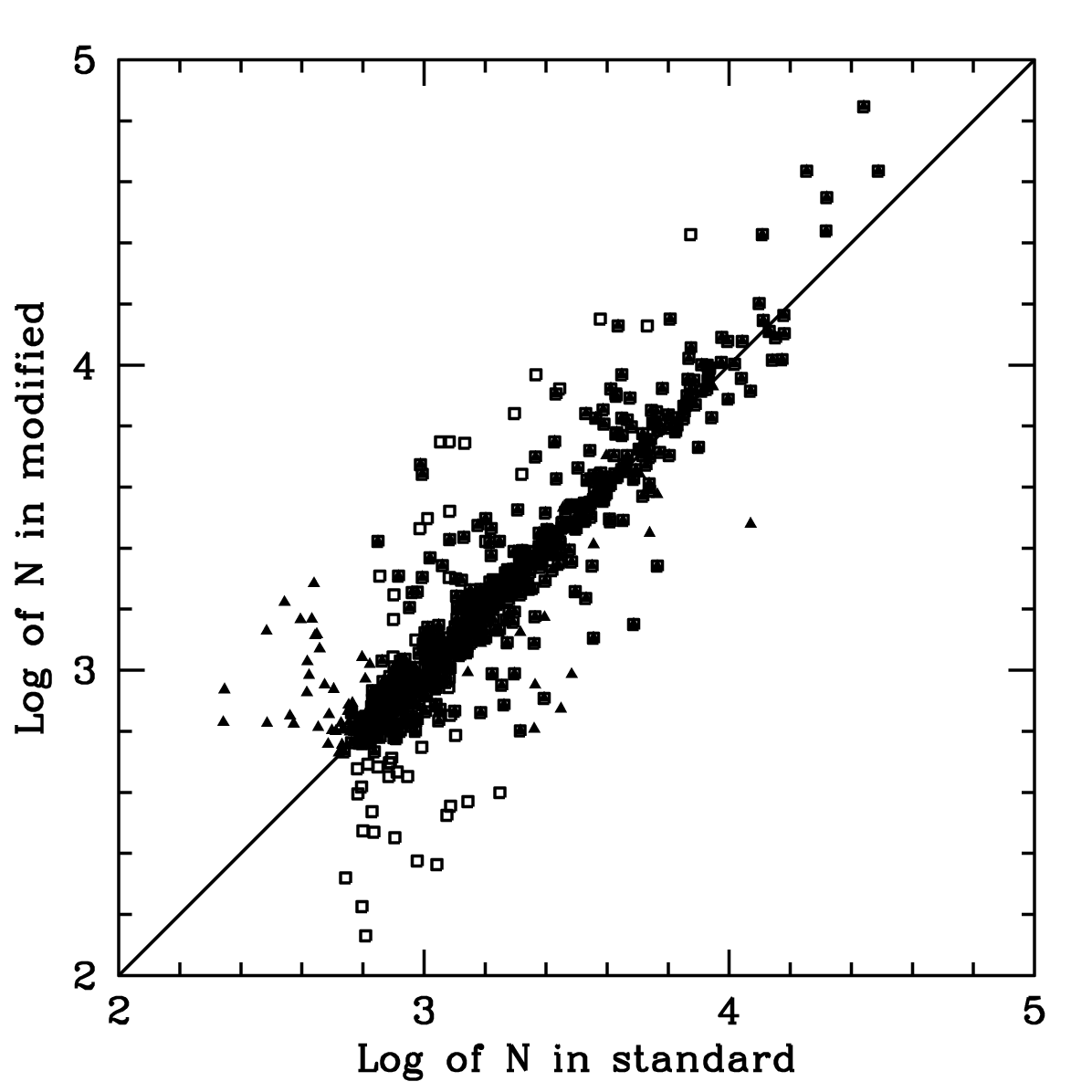}
   \caption{The panel on the top is for $\alpha=-1.0$ vs. standard, the bottom is $\alpha=0.5$ vs. standard. The points come from selecting big haloes in each realization of the modified gravity simulation and then finding the corresponding halo in the standard gravity simulation, and then the reverse. N is the number of particles in the identified halos. The tilt shows that in stronger gravity massive halos are likely to have more particles. The ``tails'' in the lower left are unimportant.}
   \label{fig:nmod}
\end{figure}

Because the simulations all had the same initial phases, we were able to perform a slightly more stringent test.  Namely, we directly compared the masses of individual halos in the $\alpha=0.5$ and $\alpha=-1$ runs with those when $\alpha=0$ (i.e. standard gravity).  The filled triangles in Figure~\ref{fig:nmod} show the result of selecting the most massive halos in the $\alpha\ne 0$ runs and plotting the number of particles they contain versus the number of particles they contained in the $\alpha=0$ run.  The open squares show the result of making the selection in the $\alpha=0$ run.  The top and bottom panels show results for $\alpha = -1$ and $0.5$ respectively.  This illustrates that with stronger gravity (larger $\alpha$), a given halo is more likely to become more massive, but this only matters for halos more massive than about $\bar\rho r_s^3$.   

\section{Conclusions}
We presented a study of nonlinear effects in a model with a modified gravitational potential (equation~\ref{potentialeq}).  In particular, we showed how the spherical evolution model is modified, and the effect this has on the abundance of virialized objects.  Halos are more massive in models where gravity is stronger on large scales (Figure~\ref{fig:nmod}), although this effect is only important for sufficiently massive objects whose evolution brings them close to the scale $r_s$ on which gravity was modified.  The effect this has on the abundance of massive objects depends on how the models are normalized.  If normalized so that the fluctuation field is the same at early-times (CMB-normalized), then the models with $\alpha>0$ have more massive halos (solid curves in Fig.~\ref{fig:comparisonlinear} differ from unity).  This remains true, although the dependence on $\alpha$ is reduced, if the models are normalized to have the same ($\alpha$-dependent) linear theory rms fluctuations today (compare solid and dashed curves in Fig.~\ref{fig:comparisonlinear}).  If normalized to have the same (linear theory) rms fluctuations today, whatever the value of $\alpha$, then the trends can be reversed (compare dotted curves with unity in Fig.~\ref{fig:comparisonlinear}).  This last normalization convention is sometimes known as `cluster-normalized':  our work suggests that, in the context of modified gravity models, this jargon is misleading!

We showed that our analysis captures the essence of the trends seen in the simulations (Figure~\ref{fig:compareall}), so, in principle, the abundance of rich clusters should place interesting constraints on modifications to the gravitational potential.  In particular, the modification to cluster abundances cannot be reproduced by standard gravity with initial conditions modified to match the change in the linear theory power spectrum; the differences in abundance can be larger than ten percent for sufficiently massive halos (Figure~\ref{fig:comparisonlinear}).  However, to use cluster counts quantitatively in this way, our analysis should be extended to models in which objects form from an ellipsoidal collapse, as was necessary for standard gravity \cite{st99,smt01}.  

Our analysis also helps to understand an interesting fact about the shape of the nonlinear power spectrum in modified gravity theories.  Figure~7 in \cite{fritz} shows that whereas the large scale power spectrum in modified theories may be rather different than in standard gravity (because the linear growth factor is modified), the power on small scales ($k>1$) is unchanged.  Our analysis shows that, because small mass objects were never larger than the scale $r_s$ on which gravity was modified, they are not affected by the modification, so their abundance is the same as in the standard case. This is not affected by the initial conditions that we choose so that in both when $\alpha$ is non-zero, and when the power spectrum is changed so that we end up with a final power spectrum the same as in the case of modified gravity, the abundances of small halos is unaffected.  In addition, their formation histories will also be unchanged, so their internal structural parameters (shapes, density profiles) are also unchanged.  In the halo model description of large scale structure \cite{cooraySheth}, the power at $k>1$ is dominated by small mass halos.  Since these are the ones for which gravity is essentially unmodified, the small-scale power spectrum is also unchanged.  

\begin{figure*}[tp]
   \centering
   \includegraphics[width=0.45\hsize]{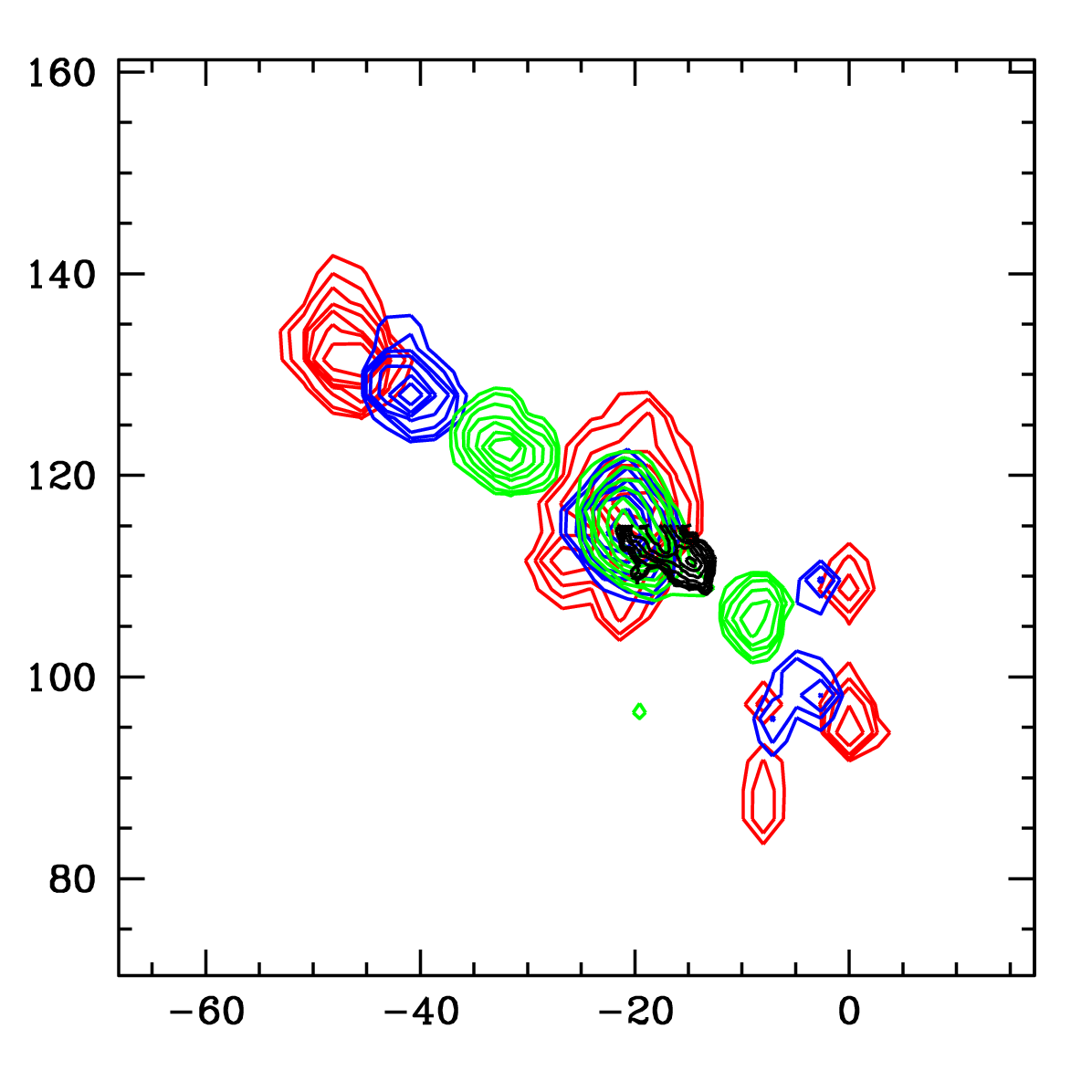}
   \includegraphics[width=0.45\hsize]{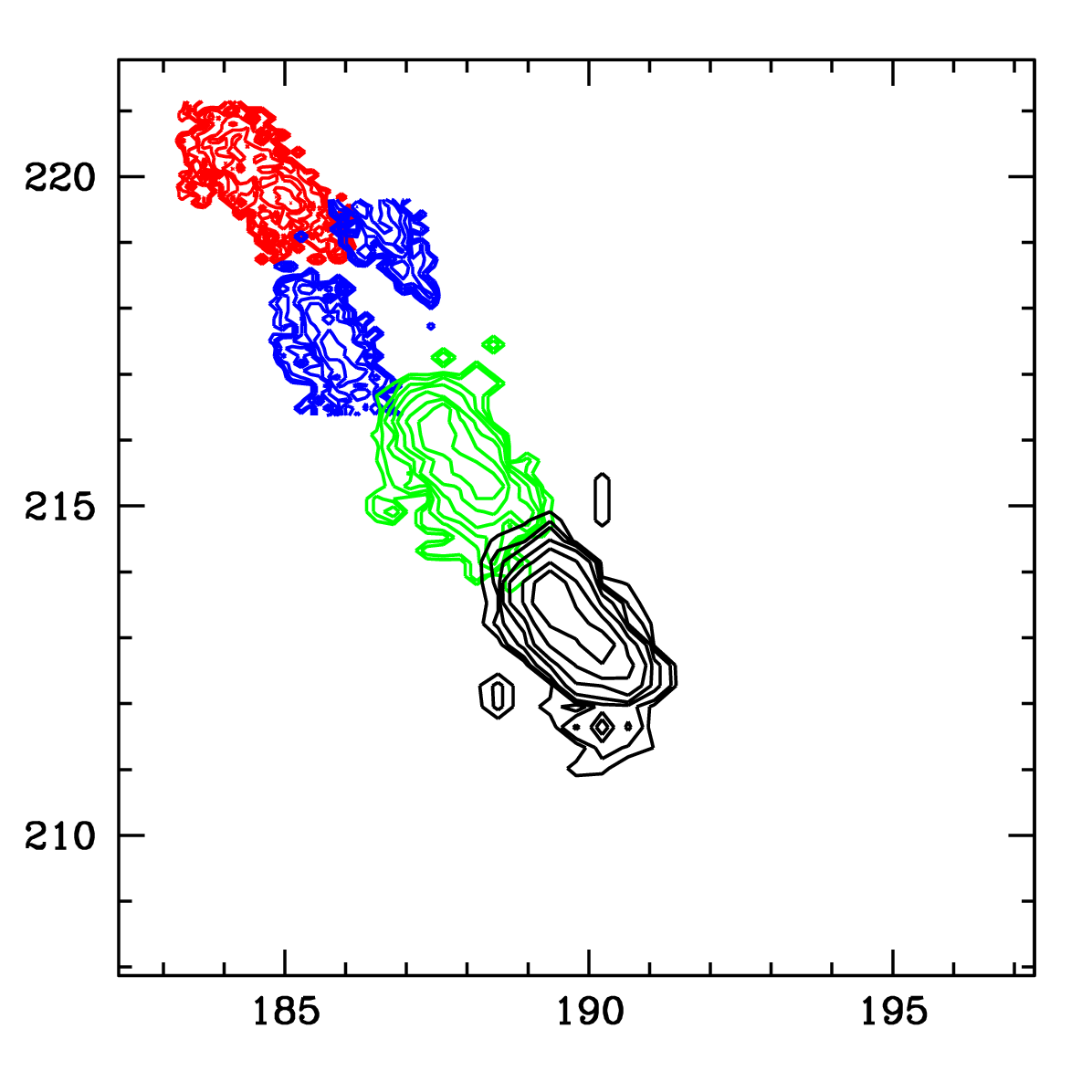}
   \caption{(Color online) Spatial distribution of a massive halo chosen from the $\alpha=0.5$ (left) and $-1$ (right) simulations. The black, green, blue, and red contours show the positions of the particles that make up the chosen halo in the $\alpha = 0.5, 0, -0.5$, and $-1$ simulations, respectively. In the left panel, note that the central dense halo spreads out for weakening gravity, and in the right panel the upper left halo merely moves.  Axes show the $(x,y)$ coordinates of the halos in grid coordinates, $100/256 h^{-1}$Mpc.}
   \label{fig:contours}
\end{figure*}

Figure~\ref{fig:contours} shows the result of a complementary study.  In this case, we selected a massive halo from one of the simulations (say $\alpha=0.5$), and then looked at where its particles were in the other runs (with different $\alpha$).  The figure illustrates clearly that when $\alpha=0.5$, then the particle distribution is more compact.  E.g., in the top panel, the large halo in the $\alpha=0.5$ run is broken up into three smaller haloes in the $\alpha = -1$ run.  The bottom panel shows another effect:  that the particles which made up a halo in the $\alpha=-1$ run are in a different location in the $\alpha=0.5$ run, suggesting that the peculiar velocities of the most massive halos are also sensitive to $\alpha$.  

The sequence of contours associated with the different $\alpha$ runs look rather similar to the time evolution of an object in, say, an $\alpha=0$ model.  Thus, the most massive halos in models with $\alpha<0$ may be like high-redshift versions of the most massive halos in models with $\alpha>0$.  Therefore, these figures suggest that the galaxy populations in massive clusters may be rather different in models with large $\alpha$ than when $\alpha$ is small.  In particular, it is likely that the difference between the cluster and field galaxy populations increases as the strength of gravity on large scales increases.  This is largely a consequence of the different $\sigma_8$ values in these runs -- so cluster $M/L$ ratios, currently used to to constrain $\sigma_8$ \cite[e.g.][]{tinkerML,flatLsat}, may one day be used to constrain modifications to gravity.  

Furthermore, in standard gravity models, there is a strong correlation between evolution and environment \cite{moWhite, st02, effectiveCosmo}.  There are two reasons to suspect that this will be different if the gravitational potential is modified.  First, in the standard model, the correlation between local environment and evolution is a consequence of Birkhoff's Theorem.  Birkhoff's Theorem is lost when the force law is modified (it is this which modified our spherical evolution model from the standard case).  And second, Figure~\ref{fig:contours} shows that the time scale for the assembly of objects is modified.  The environmental dependence of galaxy properties is in rather good agreement with the standard model \cite{drexel, abbasSheth, blantonBerlind}, so it may be that this will one day provide interesting constraints on $\alpha$.  This is the subject of work in progress.  In standard gravity, the formation and abundance of voids can be estimated using similar methods to those used for clusters \cite{voids} -- extending our analysis of the modified potential to voids is also work in progress.  

\bigskip
We would like to thank Francis Bernardeau, Antonaldo Diaferio, Lam Hui, Kyle Parfrey and Roman Scoccimarro for helpful discussions and encouragement.

\appendix
\section{Fitting formulae for the `moving' barriers}

This Appendix provides fits to the barriers in Figure \ref{fig:barriershapes}. First, let $x=\text{log} ~M/(\bar{\rho}\,r_s^3)$, where $M$ is measured in units of $10^{10}h^{-1}\,M_\odot$ and $r_s$ is measured in $h^{-1}$Mpc, then for positive $\alpha$, 
\begin{equation}
\frac{\delta_c}{\delta_c (\text{standard})} 
 = 1 -  (\alpha/26.84)   \left( 3 M/4 \pi r_s^3\right)^{0.33},
\end{equation}
whereas for negative $\alpha$,
\begin{equation}
\frac{\delta_c}{\delta_c (\text{standard})}
 = 1 +\lvert\alpha/8.38\rvert^{1.5} \left( 3M/4 \pi r_s^3\right)^{0.48}.
\end{equation}
These fits are accurate to a few percent in the range of $(-7,3)$ for $x$, $r_s$ up to $20h^{-1}$ Mpc, and $\alpha$ between $-1$ and 1. For $r_s=5h^{-1}$ Mpc, the high end of the range of $x$ corresponds to $10^{16}h^{-1}\,M_\odot$.  The dependence here makes sense, because we expect the mass scale of the modification to scale as $r_s^3$.  As for the $\alpha$ dependence, we can see from figure \ref{fig:barriershapes} that for negative $\alpha$ the deviation from standard gravity is stronger, hence the dependence on $\alpha$ for positive $\alpha$, and $|\alpha|^{1.5}$ for negative $\alpha$.

In principle, one can use these expressions to generate analytic approximations to the halo mass function as follows.  For a given initial power spectrum, $x$ can be written as a function of $S_i(M)$; this specifies the barrier shape in the units which are useful for the excursion set approach.  Then insert this barrier shape into the expressions for the first crossing distribution given by \cite{st02}.  The ratio of this first crossing distribution to that associated with a barrier of height $\delta_c$ quantifies the change in halo abundances which is due to $\alpha$ and $r_s$.  Multiplying this ratio by the actual halo abundances in the standard model \cite{st99} yields an analytic expression for the abundances in the modified model.

\label{lastpage}
\end{document}